\documentclass[]{bytedance_seed}

\usepackage[toc,page,header]{appendix}
\usepackage{calc}
\usepackage{float}

\usepackage{minitoc}
\usepackage{graphicx}
\usepackage{subcaption}
\usepackage{pifont}
\usepackage{makecell}
\usepackage{booktabs}
\usepackage{multirow}
\usepackage{float}
\usepackage{enumitem}

\usepackage{mathrsfs}
\usepackage{adjustbox}
\usepackage{multirow}
\usepackage{multirow}
\usepackage{multicol}
\usepackage{tcolorbox}
\usepackage{changepage}
\usepackage{graphicx}
\usepackage{amssymb}
\usepackage{array}
\usepackage{bm}
\usepackage{hyperref}
\usepackage{minitoc}

\newcommand{\fittowidth}[1]{%
  \sbox0{#1}%
  \ifdim\wd0>\textwidth
    \resizebox{\textwidth}{!}{\usebox0}%
  \else
    \usebox0%
  \fi
}

\title{R3-VAE: Reference Vector-Guided Rating Residual Quantization VAE for Generative Recommendation}
\author{ByteDance Toutiao Recommendation Team}

\abstract{
Generative Recommendation (GR) has gained traction for its merits of superior performance and cold-start capability. As the vital role in GR, Semantic Identifiers (SIDs) represent item semantics through discrete tokens.
However, current techniques for SID generation based on vector quantization face two main challenges: (i) training instability, stemming from insufficient gradient propagation through the straight-through estimator and sensitivity to initialization; and (ii) inefficient SID quality assessment, where industrial practice still depends on costly GR training and A/B testing.
To address these challenges, we propose \textbf{Reference Vector-Guided Rating Residual Quantization VAE ($\textit{R}^\textit{3}$\textit{-VAE})}. This framework incorporates three key innovations: (i) a reference vector that functions as a semantic anchor for the initial features, thereby mitigating sensitivity to initialization; (ii) a dot product-based rating mechanism designed to stabilize the training process and prevent codebook collapse; and (iii) two SID evaluation metrics, Semantic Cohesion and Preference Discrimination, serving as regularization terms during training.
Empirical results on six benchmarks demonstrate that $\textit{R}^\textit{3}$\textit{-VAE} outperforms state-of-the-art methods, achieving an average improvement of 14.5\% in Recall@10 and 15.5\% in NDCG@10 across three public datasets (Beauty, Sports, and Toys). 
Furthermore, we perform GR training and online A/B tests on Toutiao. Our method achieves a 1.62\% improvement in MRR and a 0.83\% gain in StayTime/U versus baselines.
Additionally, we employ $\textit{R}^\textit{3}$\textit{-VAE} to replace the
item ID of CTR model, resulting in significant improvements in content cold start by $15.36\%$, corroborating the strong applicability and business value in industry-scale recommendation scenarios. 
Crucially, our proposed metrics exhibit a stronger correlation with downstream performance, evidenced by Spearman's rank correlation coefficients of 0.94 and 0.90 in the GR and CTR experiments.
}
\checkdata[Github Repository]{\url{https://github.com/wwqq/R3-VAE}}
\begin{document}
\maketitle

%不需要目录就注释掉 注意目录不要和第一页放在一块 要有\newpage
%\newpage
%\tableofcontents
%\newpage

\section{Introduction}
\label{sec:intro}

With the rise of Large Language Models (LLMs), there has been a paradigm shift in the field of recommender systems. 
Generative Recommendation (GR)~\cite{rajput2023recommender,hou2025generating,qu2025tokenrec,deng2025onerec,liu2025generative,fu2025forge,ju2025generative,hou2025generative,geng2022recommendation} is gradually replacing traditional ``retrieval-reranking'' pipelines.
It directly outputs item identifiers via generative models, eliminating intermediate embedding tables, ANN indexes, and reranking modules for scalable end-to-end recommendation. 
As the key enabler of GR, Semantic Identifier (SID)~\cite{hou2023learning,rajput2023recommender,deng2025onerec} discrete tokens that capture item semantic attributes, enabling an interpretable mapping to human-understandable categories and seamless integration with pre-trained language models~\cite{zheng2024adapting,zhai2024actions,liu2024multi}. 
Current SID generation relies on vector quantization (VQ), which compresses continuous embeddings into discrete codewords via a learnable codebook. Among VQ methods, Residual Quantization (RQ) is a cornerstone with implementations such as RQ-VAE~\cite{lee2022autoregressive,zeghidour2021soundstream} (in TIGER~\cite{rajput2023recommender}) and residual K-Means~\cite{luo2024qarm} (in OneRec~\cite{deng2025onerec}), which leverage hierarchical ``coarse-to-fine'' residual quantization for efficient learning. Complemented by VQ-Rec’s OPQ~\cite{hou2023learning,ge2013optimized} for codebook optimization, these advances establish VQ-based quantization as the de facto standard for SID learning in GR.

Despite significant progress, state-of-the-art SID generation techniques also exhibit limitations that impede their performance in GR systems.
A primary drawback is training instability stemming from inadequate gradient propagation during quantization. 
RQ-VAE, for instance, relies on the Straight-Through Estimator (STE) to approximate gradients across the non-differentiable codebook lookup operation ~\cite{lee2022autoregressive,zeghidour2021soundstream}. 
This approximation provides a biased gradient estimator. 
It often fails to propagate fine-grained information from the reconstruction loss effectively, leading to codebook collapse where the majority of vectors map to a small subset of codewords. 
While residual K-Means quantization ~\cite{luo2024qarm} partially mitigates this issue, it is highly affected by the initial cluster centers and tends to prolong convergence or get trapped in local optima. \looseness=-1

Another practical challenge in GR development is the lack of quantifiable metrics strongly correlated with downstream performance for SID quality assessment. 
Current approaches validate SID indirectly via downstream GR metrics (e.g., NDCG, Recall) in end-to-end evaluations.
While this reflects real-world utility, it is computationally costly. 
This inefficiency severely hinders model iteration, as quantization module modifications cannot be rapidly evaluated without full downstream validation, impeding GR system design progress. 
Moreover, existing SID evaluation metrics proposed by some methods, such as Collision Rate in~\cite{li2025semantic} and the Gini coefficient~\cite{fu2025forge}, exhibit limited correlation with downstream tasks and are not conducive to end-to-end optimization.

To address these limitations, we propose \textit{Reference Vector-Guided Rating Residual Quantization VAE ($\textit{R}^\textit{3}$-VAE)}, a novel framework tailored for high-quality SID generation in GR systems. Our key contributions are as follows:
\textbf{(i)}
We introduce the reference vector that serves as a semantic anchor for initial item feature processing. By projecting input embeddings by this reference vector and computing the residual, we align the quantization process with the semantic structure of recommendation data from the outset. This design mitigates initialization sensitivity and ensures early residual vectors capture preference-relevant information, addressing the training instability and convergence challenges inherent in RQ-VAE and residual K-Means quantization.
\textbf{(ii)}
We replace STE with a rating mechanism that employs dot product to propagate reconstruction loss gradients. We compute ``ratings'' between codewords and residual with the dot product to measure angular similarity (e.g., semantic alignment) to quantify residual-codebook relevance. Unlike STE, this design retains fine-grained gradient information across quantization steps, stabilizing training and accelerating convergence. It preserves the hierarchical advantages of RQ-VAE~\cite{lee2022autoregressive} and residual K-Means~\cite{luo2024qarm} while eliminating codebook collapse, addressing the core instability of existing methods. 
\textbf{(iii)}
We propose two quantifiable and optimizable metrics for SID evaluation: Semantic Cohesion (SC) and Preference Discrimination (PD). Specifically, SC measures the intra-cluster semantic consistency by computing the average cosine similarity between items in the same SID. Higher SC indicates stronger preference alignment within clusters. PD quantifies the inter-cluster divergence by calculating the average cosine distance between SIDs from different clusters and their respective centroids. Lower PD indicates distinct preference patterns across clusters.
% These metrics exhibit strong correlation (Spearman's \(\rho = 0.90\)) with downstream performance, enabling efficient SID quality assessment.
Both metrics exhibit a stronger correlation with downstream performance than the existing measures, including collision rate and Gini coefficient. This superior performance is substantiated by our online A/B test, where we observed a Spearman’s rank correlation coefficient of $\rho = 0.90$.

Empirical results on six benchmarks demonstrate that $\textit{R}^\textit{3}$\textit{-VAE} outperforms state-of-the-art methods, achieving an average improvement of 14.5\% in Recall@10 and 15.5\% in NDCG@10 across three public datasets (Beauty, Sports, and Toys).
More importantly, we conduct comprehensive experiments on a large-scale industrial dataset derived from Toutiao, comparing $\textit{R}^\textit{3}$\textit{-VAE} with mainstream industrial baselines under the OneRec framework. Offline evaluation results show $\textit{R}^\textit{3}$\textit{-VAE} achieves an MRR of 0.628, representing 3.80\% and 1.62\% improvements over RQ-VAE and residual K-Means, respectively. Online A/B test further confirms its practical value. $\textit{R}^\textit{3}$\textit{-VAE} improves StayTime/U (defined as average time spent by a user per day) by 1.59\% compared to RQ-VAE and by 0.82\% compared to residual K-Means.
Additionally, it improves LongStay/U (the count of documents on which a user’s stay time exceeds ten seconds per user) by 0.60\% in comparison to RQ-VAE and 0.28\% to residual K-Means. 
Inspired by the performance observed in GR, we further employed the SID to replace the item ID in traditional CTR models, resulting in significant improvements in cold start and user engagement metrics. Notably, $\textit{R}^\textit{3}$\textit{-VAE} facilitates a substantial increase of 15.36\% in the total click volume of cold-start content (defined as content with impression counts below 512) and achieves a 0.65\% enhancement in stay duration per user when compared to R-KMeans. 
These findings underscore the excellent applicability and business value of $\textit{R}^\textit{3}$\textit{-VAE} in real-world industrial recommendations.

The remainder of this paper is structured as follows: Section 2 reviews related work on GR and SID generation. Section 3 details the architecture and optimization of $\textit{R}^\textit{3}$\textit{-VAE}. Section 4 presents experimental results. Section 5 concludes the paper.
\section{Related Work}
\label{sec:related_work}
\subsection{Generative Recommendation}
Generative Recommendation (GR)~\cite{rajput2023recommender,hou2025generating,qu2025tokenrec,deng2025onerec,liu2025generative,fu2025forge,ju2025generative,ji2024genrec,wang2023generative,liu2025onerec} has emerged as a transformative paradigm in recommender systems, deviating from traditional multi-stage ``retrieval-reranking'' pipelines by leveraging generative models to directly output target item identifiers (e.g., semantic IDs) for end-to-end recommendation without relying on intermediate embedding tables, ANN indexes, or separate reranking modules. 

This paradigm shift is driven by the powerful sequence modeling and semantic understanding capabilities of Large Language Models (LLMs) and advanced generative architectures, addressing long-standing limitations of traditional recommendation systems. 
OneRec~\cite{deng2025onerec} and TIGER~\cite{rajput2023recommender} leverage vector quantization to map items into semantic ID sequences, enabling generative models to capture fine-grained user preferences and item correlations beyond the scope of collaborative filtering. Recent advancements further enhance GR with explicit reasoning capabilities (e.g., OneRec-Think~\cite{liu2025onerec}) and adaptive domain alignment (e.g., LGSID~\cite{jiang2025llm}), allowing the framework to address scenario-specific constraints by incorporating task-specific signals into the generation process.

\subsection{Semantic Identifier}
VQ-Rec~\cite{hou2023learning} proposed an optimized product quantization (OPQ)~\cite{ge2013optimized} based method to discretize encodings of items to obtain semantically-rich and distinguishable item codes.
CCFRec~\cite{liu2025bridging} used PQ~\cite{jegou2010product} and RQ to map item embeddings into multi-level discrete codes.
UniSearch~\cite{chen2025unisearch} set VQ-VAE as the tokenizer to discretize multi-modal information into SID.
TIGER~\cite{rajput2023recommender}, LETTER~\cite{wang2024learnable}, SETRec~\cite{lin2025order}, GNPR-SID~\cite{wang2025generative}, EGA-V2~\cite{zheng2025ega}, DAS~\cite{ye2025dual}, UIST~\cite{liu2024discrete}, CoST~\cite{zhu2024cost}, and MM-RQ-VAE~\cite{wang2025empowering} adopted a hierarchical quantizer called Residual Quantized VAE (RQ-VAE)~\cite{lee2022autoregressive,zeghidour2021soundstream} on content embeddings to generate tokens that form the semantic Identifier. 
SEATER~\cite{si2024generative} used hierarchical constrained K-Means~\cite{bradley2000constrained} to generate equal-length identifiers.
OneRec~\cite{deng2025onerec} introduced a multi-level balanced quantitative mechanism with the residual K-Means quantization algorithm~\cite{luo2024qarm}. 
Besides, RPG~\cite{hou2025generating}, TokenRec~\cite{qu2025tokenrec}, ETEGRec~\cite{liu2025generative}, and so on~\cite{tan2025pcr,fu2025forge} have proposed several improved schemes for generating Semantic Identifier (SID). 

However, these SID generation approaches have some obvious drawbacks, such as unstable training, dependence on initialization, slow convergence, and insufficient expressive ability, which lead to suboptimal performance. In this paper, we propose $\textit{R}^\textit{3}$\textit{-VAE} to address the above issues. It has achieved significant improvements in downstream generative recommendation tasks, outperforming previous SID generation methods.

\section{Method}

\begin{figure*}
  \centering
   \includegraphics[width=\hsize]{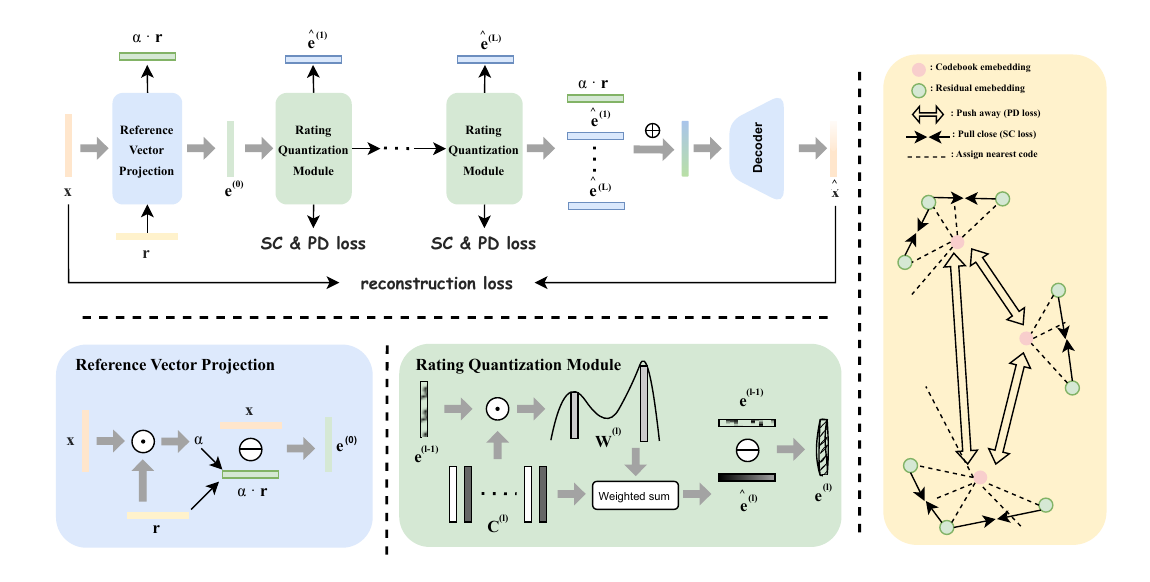}
   \caption{
  The overall pipeline of $\textit{R}^\textit{3}$\textit{-VAE}. The framework takes item continuous embeddings $\mathbf{x}$ as input, processes them via the reference vector projection layer, hierarchical rating quantization layers, decoder, and outputs reconstructed embeddings. The reconstruction loss and proposed SC \& PD loss jointly optimize the model. SC loss pulls the residual embeddings in the same cluster close, and PD loss pushes the codewords embedding away.}
  \label{fig:pipeline}
\end{figure*}

This section details the architecture and optimization of $\textit{R}^\textit{3}$\textit{-VAE}, a framework designed to address the limitations of existing SID generation methods (training instability, inefficient evaluation) for Generative Recommendation (GR). The $\textit{R}^\textit{3}$\textit{-VAE} consists of three core components: a Reference Vector Projection Layer for semantic anchoring, a Dot Product-Based Rating Quantization Module for gradient-preserving residual updates, and a pair of SID Quality Metrics (SC/PD) for direct evaluation. Figure~\ref{fig:pipeline} provides an overview of the end-to-end pipeline.

\subsection{Architecture Overview}
$\textit{R}^\textit{3}$\textit{-VAE} processes item continuous embeddings into high-quality SIDs through a hierarchical residual quantization workflow, integrating the reference vector and rating mechanism into a Variational Autoencoder (VAE) backbone. The pipeline follows three key steps:
(i) Reference Vector Projection: Input embeddings are projected by a learnable reference vector to generate semantically aligned initial residuals.
(ii) Hierarchical Rating Quantization: Residuals are iteratively quantized via a dot product-based rating mechanism, preserving gradient flow to avoid codebook collapse.
(iii) SID Generation \& Evaluation: Quantized codewords are concatenated to form SID, whose quality is directly assessed using SC and PD before downstream GR deployment.

\subsection{Reference Vector Projection Layer}
The goal of this layer is to align the initial residual computation with the semantic structure of recommendation data, mitigating the initialization sensitivity of RQ-VAE and residual K-Means. 

\subsubsection{Reference Vector Definition}
We introduce a learnable reference vector $\mathbf{r} \in \mathbb{R}^d$, where $d$ denotes the dimension of input item embeddings $\mathbf{x} \in \mathbb{R}^d$. The vector $\mathbf{r}$ is optimized end-to-end to capture the semantic center of the recommendation dataset—e.g., aggregating preference patterns of dominant semantic attributes of core items.

\subsubsection{Projection and Initial Residual Calculation}
For an input embedding $\mathbf{x}$, we first compute its semantic projection onto $\mathbf{r}$ using the dot product (to measure angular similarity, aligned with recommendation relevance). The projection scalar $\alpha$ and initial residual $\mathbf{e}^{(0)}$ are defined as:  
\begin{equation}
    \alpha = \frac{\mathbf{x} \cdot \mathbf{r}}{\|\mathbf{r}\|^2} \tag{1}
\end{equation}
\begin{equation}
    \mathbf{e}^{(0)} = \mathbf{x} - \alpha \cdot \mathbf{r} \tag{2}
\end{equation}
where $\alpha$ quantifies the semantic affinity between $\mathbf{x}$ and $\mathbf{r}$, and $\mathbf{e}^{(0)}$ represents the component of $\mathbf{x}$ that deviates from the semantic center. This ensures $\mathbf{e}^{(0)}$ retains preference-relevant information, unlike arbitrary residuals in RQ-VAE.

\subsection{Rating Quantization Module}
This layer replaces the STE-based gradient approximation in RQ-VAE with a dot product rating mechanism, enabling continuous gradient propagation across quantization steps. We adopt a hierarchical design with $L$ quantization layers (consistent with RQ’s ``coarse-to-fine'' paradigm), where each layer $l$ ($1 \leq l \leq L$) processes the residual from the previous layer and outputs a codeword.

\subsubsection{Codebook Initialization}
Each quantization layer $l$ maintains a codebook $\mathbf{C}^l = \{\mathbf{c}^l_1, \mathbf{c}^l_2, ..., \mathbf{c}^l_M\} \in \mathbb{R}^{d \times M}$, where $M$ is the number of codewords per layer (fixed across layers for simplicity). $\textit{R}^\textit{3}$\textit{-VAE} initializes $\mathbf{C}^l$ via semantic clustering: we cluster the initial residuals $\mathbf{e}^{(0)}$ of the training set using K-Means, then set $\mathbf{C}^1$ to the cluster centroids. Subsequent codebooks $\mathbf{C}^l$ ($l>1$) are initialized using centroids of residuals from layer $l-1$, ensuring alignment with hierarchical semantic structure. We also ablate the impact of initialization in subsection~\ref{sec:train_stable}.

\subsubsection{Dot Product Rating Calculation}
For the residual $\mathbf{e}^{(l-1)}$ from layer $l-1$, we compute a rating score $s^l_k$ between $\mathbf{e}^{(l-1)}$ and each codeword $\mathbf{c}^l_k \in \mathbf{C}^l$ using the normalized dot product. The dot product is chosen to measure angular similarity (semantic alignment) rather than Euclidean distance (geometric proximity), which better matches recommendation relevance:  
\begin{equation}
    s^l_k = \frac{\mathbf{e}^{(l-1)} \cdot \mathbf{c}^l_k}{\|\mathbf{e}^{(l-1)}\|_2 \times \|\mathbf{c}^l_k\|_2} \tag{3}
\end{equation}
To normalize scores into a probability distribution, we apply the softmax function to obtain the rating weight $w^l_k$ for each codeword:  
\begin{equation}
    w^l_k = \frac{\exp(s^l_k)}{\sum_{k'=1}^M \exp(s^l_{k'})} \tag{4}
\end{equation}
Here, $w^l_k$ represents the confidence that $\mathbf{c}^l_k$ is the optimal codeword for $\mathbf{e}^{(l-1)}$. Unlike RQ-VAE’s hard codebook lookup (non-differentiable), the soft $w^l_k$ enables continuous gradient flow.

\subsubsection{Residual Update and Codeword Selection}
The quantized representation $\hat{\mathbf{e}}^{(l)}$ at layer $l$ is a weighted sum of codewords (weighted by $w^l_k$), and the residual for the next layer $\mathbf{e}^{(l)}$ is updated as:  
\begin{equation}
    \hat{\mathbf{e}}^{(l)} = \sum_{k=1}^K w^l_k \cdot \mathbf{c}^l_k \tag{5}
\end{equation}
\begin{equation}
    \mathbf{e}^{(l)} = \mathbf{e}^{(l-1)} - \hat{\mathbf{e}}^{(l)} \tag{6}
\end{equation} 
Here, $K$ is the top-$K$ weights.
After processing all $L$ layers, the SID for input $\mathbf{x}$ is formed by concatenating the indices of the top-weighted codeword from each layer:  
\begin{equation}
    \text{SID}(\mathbf{x}) = [\arg\max_{k} w^1_k, \arg\max_{k} w^2_k, ..., \arg\max_{k} w^L_k] \tag{7}
\end{equation}
% This hierarchical SID balances compactness (length $L$) and expressiveness (total codewords $K^L$).

\subsection{SID Quality Metrics}
To bypass the computationally costly downstream GR validation step, we introduce two direct, quantifiable metrics for assessing SID quality: Semantic Cohesion (SC), which measures intra-cluster semantic consistency, and Preference Discrimination (PD,) which quantifies inter-cluster preference divergence. These metrics draw inspiration from prior works~\cite{li2025semantic,deng2025onerec,fu2025forge} on cluster validity assessment, adapted to align with the semantic and preference-oriented nature of SID in generative recommendation scenarios.

\subsubsection{Semantic Cohesion (SC)}
SC measures the consistency of semantics by calculating the average cosine similarity between the quantitative embeddings of items and their corresponding code embeddings within the same cluster. For a cluster $\mathcal{G}$ (containing a set of items), let $\mathbf{q}_i$ denote the quantitative embedding of item $i \in \mathcal{G}$, and $\mathbf{q}_j$ denote the quantitative embedding of another item $j \in \mathcal{G}$. The SC for cluster $\mathcal{G}$ is defined as:  
\begin{equation}
    \text{SC}(\mathcal{G}) = \frac{2}{|\mathcal{G}|^2-|\mathcal{G}|} \sum_{i,j \in \mathcal{G}, \, i \neq j} \frac{\mathbf{q}_i \cdot \mathbf{q}_j}{\|\mathbf{q}_i\| \cdot \|\mathbf{q}_j\|} \tag{8}
\end{equation}
where $|\mathcal{G}|$ is the number of items in cluster $\mathcal{G}$, $\cdot$ denotes the dot product operation, and $\|\cdot\|$ represents the L2 norm of a vector. The overall SC of the model is the average of $\text{SC}(\mathcal{G})$ across all clusters $\mathcal{C}$. 
\begin{equation}
\label{equ:9}
    \text{SC} = \frac{1}{|\mathcal{C}|} \sum_{\mathcal{G} \in \mathcal{C}} \text{SC}(\mathcal{G}) \tag{9}
\end{equation}
A higher SC value indicates that the quantitative embeddings and code embeddings of items within the same cluster are more consistent.\looseness=-1

\subsubsection{Preference Discrimination (PD)}
PD quantifies how distinct the preference patterns of different SID clusters are. For two disjoint clusters $\mathcal{G}_a$ and $\mathcal{G}_b$, their preference divergence is measured by the distance between their centroid vectors $\bar{\mathbf{p}}_{\mathcal{G}_a}$ and $\bar{\mathbf{p}}_{\mathcal{G}_b}$. The overall PD is the average distance across all pairs of clusters:  
\begin{equation}
\label{eqa:10}
    \text{PD} = \log\left( \frac{2}{M(M - 1)}  \sum_{a=1}^{M} \sum_{\substack{b=a+1}}^{M} e^{-t \cdot (1 - \frac{\vec{\mathbf{p}}_{g_a} \cdot \vec{\mathbf{p}}_{g_b}}{\|\vec{\mathbf{p}}_{g_a}\| \cdot \|\vec{\mathbf{p}}_{g_b}\|})} \right) \tag{10}
\end{equation} 
where $M$ is the total number of SID clusters. Here, we set $t$ to 2. Lower PD indicates clusters have more distinct preference patterns, reducing redundancy in downstream GR.

```latex
\begin{table*}
    \centering
    \caption{Downstream GR performance of $\textit{R}^\textit{3}$\textit{-VAE} and baselines on Beauty, Sports, and Toys. The \textbf{bold} values indicate the best performance, and the \underline{underlined} values denote the second-best performance.}
    \begin{tabular}{c|c|c|c|c|c|c|c|c|c}
    \hline
    
    \hline
    
    \hline
         \multicolumn{1}{c}{Dataset} & 
        \multicolumn{1}{c}{Method} & 
        \multicolumn{3}{c}{Recall $\uparrow$} & 
        \multicolumn{3}{c}{NDCG $\uparrow$} & 
        \multicolumn{1}{c}{Semantic $\uparrow$} & 
        \multicolumn{1}{c}{Preference $\downarrow$} \\
        \cline{3-5} \cline{6-8}
        \multicolumn{1}{c}{} & 
        \multicolumn{1}{c}{} & 
        \multicolumn{1}{c}{@10} & 
        \multicolumn{1}{c}{@20} & 
        \multicolumn{1}{c}{@30} & 
        \multicolumn{1}{c}{@10} & 
        \multicolumn{1}{c}{@20} & 
        \multicolumn{1}{c}{@30} & 
        \multicolumn{1}{c}{Cohesion} & 
        \multicolumn{1}{c}{Discrimination} \\
        \hline
        Beauty&VQ-VAE&0.055&0.072&0.086&0.031&0.036&0.039&0.85&-0.17 \\
        &RQ-VAE&0.059&0.091&0.111&0.033&0.041&0.047&0.93&-1.18 \\
        &KMeans&0.060&0.095&0.115&0.035&0.043&0.048&0.93&-0.22 \\
        &OPQ-KMeans&0.059&0.092&0.113&0.033&0.042&0.048&0.94&-1.14 \\
        &R-KMeans&\underline{0.064}&\underline{0.097}&\underline{0.121}&\underline{0.035}&\underline{0.043}&\underline{0.048}&\underline{0.96}&\underline{-1.39} \\
        &MQ&0.056&0.081&0.092&0.034&0.037&0.042&0.90&-1.03 \\
        &$\textbf{R}^\textbf{3}$\textbf{-VAE}&\textbf{0.072}&\textbf{0.102}&\textbf{0.125}&\textbf{0.039}&\textbf{0.047}&\textbf{0.052}&\textbf{0.97}&\textbf{-1.81} \\
        \hline
        
    \hline
    
    \hline
        Sports
        &VQ-VAE&0.028&0.039&0.051&0.015&0.018&0.020&0.90&-0.11 \\
        &RQ-VAE&0.031&0.048&0.062&0.017&0.021&0.024&0.94&-1.17 \\
        &KMeans&0.033&0.054&0.069&0.018&0.023&0.026&0.91&-0.21 \\
        &OPQ-KMeans&0.030&0.044&0.057&0.015&0.019&0.023&0.90&-1.14 \\
        &R-KMeans&0.035&\underline{0.058}&\underline{0.071}&0.019&\underline{0.025}&\underline{0.028}&\underline{0.95}&\underline{-1.38} \\
        &MQ&\underline{0.036}&0.058&0.070&\underline{0.020}&0.025&0.027&\underline{0.95}&-1.16 \\
        &$\textbf{R}^\textbf{3}$\textbf{-VAE}&\textbf{0.041}&\textbf{0.063}&\textbf{0.078}&\textbf{0.022}&\textbf{0.027}&\textbf{0.030}&\textbf{0.97}&\textbf{-1.80} \\
        \hline
        
    \hline
    
    \hline
        Toys
        &VQ-VAE&0.044&0.051&0.062&0.025&0.027&0.029&0.91&-0.15 \\
        &RQ-VAE&0.051&0.074&0.089&0.028&0.035&0.038&0.94&-1.19 \\
        &KMeans&0.052&0.077&0.097&0.029&0.035&0.040&0.92&-0.22 \\
        &OPQ-KMeans&0.051&0.071&0.080&0.028&0.033&0.037&0.94&-1.15 \\
        &R-KMeans&\underline{0.058}&\underline{0.086}&\underline{0.107}&\underline{0.031}&\underline{0.038}&\underline{0.043}&\underline{0.96}&\underline{-1.39} \\
        &MQ&0.054&0.081&0.100&0.029&0.036&0.040&0.89&-1.02 \\
        &$\textbf{R}^\textbf{3}$\textbf{-VAE}&\textbf{0.066}&\textbf{0.094}&\textbf{0.114}&\textbf{0.037}&\textbf{0.044}&\textbf{0.048}&\textbf{0.98}&\textbf{-1.83} \\
    \hline
    
    \hline
    
    \hline
    \end{tabular}
    \label{tab:main_results}
\end{table*}
```

\subsection{Model Optimization}
$\textit{R}^\textit{3}$\textit{-VAE} is optimized via a hybrid loss function that combines VAE reconstruction loss (for embedding fidelity) and metric-aware regularization (for SID quality).

\subsubsection{Reconstruction Loss}
The VAE decoder reconstructs the input embedding $\mathbf{x}$ from the final residual $\mathbf{e}^{(L)}$ and the sum of all quantized representations $\sum_{l=1}^L \hat{\mathbf{e}}^{(l)}$. The reconstruction loss is the Mean Squared Error (MSE) between $\mathbf{x}$ and its reconstruction $\hat{\mathbf{x}}$:  
\begin{equation}
    \mathcal{L}_{\text{rec}} = \|\mathbf{x} - \hat{\mathbf{x}}\|^2 = \left\|\mathbf{x} - \left(\alpha \cdot \mathbf{r} + \sum_{l=1}^L \hat{\mathbf{e}}^{(l)}\right)\right\|^2 \tag{11}
\end{equation}
\subsubsection{Metric-Aware Regularization}
To align the model with SC and PD, we add a regularization term that maximizes SC and PD. Let $\text{SC}_{\text{avg}}$ denote the average SC across clusters, and $\text{PD}_{\text{avg}}$ denote the average PD across clusters. The regularization loss is:  
\begin{equation}
    \mathcal{L}_{\text{metric}} = -\text{SC}_{\text{avg}} + \text{PD}_{\text{avg}} \tag{12}
\end{equation}  

\subsubsection{Total Loss}
The total loss for $\textit{R}^\textit{3}$\textit{-VAE} is the sum of the reconstruction loss and metric regularization:  
\begin{equation}
    \mathcal{L}_{\text{total}} = \mathcal{L}_{\text{rec}} + \lambda \cdot \mathcal{L}_{\text{metric}} \tag{13}
\label{eqa:reg}
\end{equation} 
where $\lambda > 0$ is a hyperparameter balancing the two terms (set to 0.01 in experiments via cross-validation). The model is optimized using AdamW optimizer with a learning rate of $5 \times 10^{-4}$ and weight decay of $1 \times 10^{-5}$.

\section{Experiments}
To validate the effectiveness of $\textit{R}^\textit{3}$\textit{-VAE} in generating high-quality SID for Generative Recommendation (GR), we design experiments addressing three key questions: (1) Does $\textit{R}^\textit{3}$\textit{-VAE} outperform state-of-the-art SID generation methods in downstream GR tasks? (2) Are the proposed SID metrics (SC/PD) effective proxies for GR performance? (3) Do the core components of $\textit{R}^\textit{3}$\textit{-VAE} (reference vector, rating mechanism) contribute to performance improvements?

\subsection{Experimental Setup}
\subsubsection{Datasets}
We use six public recommendation datasets, including Beauty, Sports, Toys, and Clothing~\cite{hou2024bridging,rajput2023recommender,ni2019justifying}, LastFM, and ML1M, with diverse scales and application scenarios to ensure the generalizability of our method. 

\subsubsection{Baselines}
We compare $\textit{R}^\textit{3}$\textit{-VAE} with several mainstream VQ and RQ-based SID generation methods:
\textbf{VQ-VAE}~\cite{van2017neural}: A standard vector quantization variational autoencoder (VAE) using the straight-through estimator (STE) for gradient approximation without residual decomposition.
\textbf{RQ-VAE} and its variants~\cite{lee2022autoregressive,zeghidour2021soundstream}: Hierarchical residual quantization VAE.
\textbf{MQ}~\cite{qu2025tokenrec}: Masked vector based RQ-VAE.
\textbf{OPQ-KMeans} (Optimized Product Quantization KMeans)~\cite{ge2013optimized,hou2023learning}: VQ-based method that optimizes codebook orthogonality to reduce quantization error.
\textbf{Residual K-Means}~\cite{luo2024qarm,deng2025onerec}: RQ-based method that integrates K-Means clustering for codebook learning.

\subsubsection{Evaluation Metrics}
We evaluate from two perspectives: downstream GR performance for validating SID utility, and SID quality to validate the proposed metrics SC/PD.

\begin{table*}
    \centering
    \caption{Downstream GR performance of $\textit{R}^\textit{3}$\textit{-VAE} and baselines on LastFM, ML1M and Clothing. The \textbf{bold} values indicate the best performance, and the \underline{underlined} values denote the second-best performance.}
    \begin{tabular}{c|c|c|c|c|c|c|c|c|c}
    \hline

    \hline

    \hline
         \multicolumn{1}{c}{Dataset} & 
        \multicolumn{1}{c}{Method} & 
        \multicolumn{3}{c}{Recall $\uparrow$} & 
        \multicolumn{3}{c}{NDCG $\uparrow$} & 
        \multicolumn{1}{c}{Semantic $\uparrow$} & 
        \multicolumn{1}{c}{Preference $\downarrow$} \\
        \cline{3-4} \cline{5-6} \cline{7-8}
        \multicolumn{1}{c}{} & 
        \multicolumn{1}{c}{} & 
        \multicolumn{1}{c}{@10} & 
        \multicolumn{1}{c}{@20} & 
        \multicolumn{1}{c}{@30} & 
        \multicolumn{1}{c}{@10} & 
        \multicolumn{1}{c}{@20} & 
        \multicolumn{1}{c}{@30} & 
        \multicolumn{1}{c}{Cohesion} & 
        \multicolumn{1}{c}{Discrimination} \\
        \hline
        LastFM&VQ-VAE&0.042&0.088&0.118&0.018&0.030&0.036&0.40&-1.56 \\  
        &RQ-VAE&0.047&0.087&0.119&0.021&0.031&0.038&0.72&-1.65\\
        &KMeans&0.051&0.084&0.118&0.024&0.033&0.040&0.79&-1.73 \\
        & OPQ-KMeans&\underline{0.060}&\underline{0.104}&\underline{0.140}&\underline{0.030}&\underline{0.041}&\underline{0.047}&\underline{0.91}&\underline{-1.85} \\
        &R-KMeans&0.056&0.097&0.127&0.026&0.038&0.044&0.85&-1.77 \\ &MQ&0.053&0.094&0.125&0.025&0.035&0.042&0.80&-1.71 \\ %~\cite{qu2025tokenrec}
        &$\textbf{R}^\textbf{3}$\textbf{-VAE}&\textbf{0.076}&\textbf{0.117}&\textbf{0.149}&\textbf{0.037}&\textbf{0.047}&\textbf{0.054}&\textbf{0.98}&\textbf{-1.98} \\ 
        \hline
    
    \hline

    \hline
        ML1M
        &VQ-VAE&0.069&0.116&0.154&0.033&0.045&0.053&0.56&-1.69 \\           
        &RQ-VAE&0.092&0.163&0.210&0.051&0.065&0.071&0.71&-1.73  \\     
        &KMeans&0.072&0.135&0.184&0.034&0.053&0.061&0.57&-1.67 \\  
        &OPQ-KMeans&0.132&0.214&0.275&0.065&0.086&0.099&0.88&-1.83 \\  
        &R-KMeans&\underline{0.160}&\underline{0.244}&\underline{0.303}&\underline{0.080}&\underline{0.101}&\underline{0.114}&\underline{0.91}&\underline{-1.88} \\ 
        &MQ&0.101&0.168&0.215&0.053&0.070&0.080&0.78&-1.75 \\ 
        &$\textbf{R}^\textbf{3}$\textbf{-VAE}&\textbf{0.183}&\textbf{0.277}&\textbf{0.341}&\textbf{0.096}&\textbf{0.119}&\textbf{0.133}&\textbf{0.97}&\textbf{-1.98} \\ 
        \hline
    
    \hline

    \hline
        Clothing&VQ-VAE&0.009&0.016&0.023&0.004&0.006&0.008&0.73&-1.61 \\ 
        &RQ-VAE&0.011&0.018&0.024&0.005&0.007&0.008&0.81&-1.74 \\ 
        &KMeans&0.007&0.013&0.017&0.003&0.005&0.006&0.67&-1.55 \\
        &OPQ-KMeans&0.013&0.020&0.025&0.007&0.009&0.010&0.86&-1.85 \\
        &R-KMeans&0.014&0.021&0.026&\underline{0.008}&0.009&0.011&\underline{0.90}&-1.87 \\ 
        &MQ&\underline{0.017}&\underline{0.024}&\underline{0.029}&\textbf{0.011}&\underline{0.011}&\underline{0.013}&0.87&\underline{-1.89} \\ 
        &$\textbf{R}^\textbf{3}$\textbf{-VAE}&\textbf{0.018}&\textbf{0.026}&\textbf{0.033}&\textbf{0.011}&\textbf{0.012}&\textbf{0.014}&\textbf{0.96}&\textbf{-2.00} \\
    \hline
    
    \hline

    \hline
    \end{tabular}
    \label{tab:main_results_2}
\end{table*}

\paragraph{Downstream GR Performance Metrics}
We use a Transformer-based GR model (consistent across all methods) that takes SID as input to generate recommended item sequences. Key metrics:  
\textbf{Recall@K}: Fraction of relevant items in the top-K generated sequence (K=10, 20, 30).  
\textbf{NDCG@K}: Normalized Discounted Cumulative Gain (K=10, 20, 30); measures ranking quality of relevant items.  

\paragraph{SID Quality Metrics}
Semantic Cohesion (SC): Calculated via Eq.~\ref{equ:9}; higher values indicate more consistent preferences within clusters.  
Preference Discrimination (PD): Calculated via Eq.~\ref{eqa:10}; lower values indicate more distinct preferences across clusters.

\subsection{Comparison with state of the art}
Following~\cite{ju2025generative,qu2025tokenrec}, all methods are configured with the same hyperparameters for fairness. Table~\ref{tab:main_results} presents the downstream GR performance of $\textit{R}^\textit{3}$\textit{-VAE} and baselines on Beauty, Sports, and Toys datasets. From the table, we can observe that $\textit{R}^\textit{3}$\textit{-VAE} outperforms all baselines across all metrics and datasets consistently. For instance, on Beauty, $\textit{R}^\textit{3}$\textit{-VAE} achieves Recall@10 = 0.072 and NDCG@10 = 0.039, representing 12.5\% and 11.4\% improvements over R-KMeans (the strongest baseline in this set), respectively. On Sports, $\textit{R}^\textit{3}$\textit{-VAE} further demonstrates its effectiveness by boosting Recall@10 by 17.1\% and NDCG@10 by 15.8\% compared to R-KMeans, while on the Toys dataset, it delivers 13.8\% higher Recall@10 and 19.4\% higher NDCG@10 than the same baseline, confirming its robustness across diverse recommendation scenarios.  

To comprehensively validate $\textit{R}^\textit{3}$\textit{-VAE}’s generalization ability, Table~\ref{tab:main_results_2} extends the evaluation to three additional datasets: LastFM, ML1M, and Clothing. The results here reinforce $\textit{R}^\textit{3}$\textit{-VAE}’s superior performance. on LastFM, it achieves Recall@10 = 0.076 and NDCG@10 = 0.037, outperforming OPQ-KMeans by 26.7\% and 23.3\% in these two metrics. On ML1M, $\textit{R}^\textit{3}$\textit{-VAE} maintains advantage with 14.4\% and 20.0\% gains in Recall@10 and NDCG@10 over R-KMeans, respectively. On Clothing, $\textit{R}^\textit{3}$\textit{-VAE} outperforms MQ, further verifying that its superior performance compared with all these baselines.

\subsection{Generative Recommendation Performance on Industrial Dataset}
To validate the practical applicability of $\textit{R}^\textit{3}$\textit{-VAE} in real-world generative recommendation scenarios, we conduct experiments on a large-scale industrial dataset from interaction logs of Toutiao. Constrained by resources, we compare $\textit{R}^\textit{3}$\textit{-VAE} with two mainstream industrial generative model baselines: RQ-VAE and Residual K-Means (R-KMeans).  
\subsubsection{Implementation Details}
The overall experimental framework follows OneRec~\cite{deng2025onerec}. The codebook size is $8192\times8192\times8192$. The optimizer adopts AdamW with an initial learning rate of $2 \times 10^{-4}$, and all experiments are conducted on $64$ NVIDIA A100 GPUs. For generation configuration, the $r_{\text{DPO}}$ is set to 5\%, and the number of responses generated by beam search is 512. User historical behavior features include user profile information, real-time click sequences, real-time impression sequences, and recent click sequences.  
\subsubsection{Offline Evaluation}
We use Mean Reciprocal Rank (MRR) as the core offline evaluation metric. MRR calculates the reciprocal of the rank of the correct result in the predicted sequence for each sample and takes the average, reflecting the model's ability to rank correct results in top positions.  

Table~\ref{tab:industrial_online_results} presents the offline MRR performance of all methods. From Table, we observe that $\textit{R}^\textit{3}$\textit{-VAE} significantly outperforms both baselines. 
$\textit{R}^\textit{3}$\textit{-VAE} attains a superior MRR of 0.628, marking improvements of 3.80\% over RQ-VAE (MRR = 0.605) and 1.62\% over R-KMeans (MRR = 0.618).
This confirms that $\textit{R}^\textit{3}$\textit{-VAE}'s core innovations (reference vector-based semantic anchoring and gradient-preserving rating mechanism) effectively enhance the quality of generated SID, thereby improving the accuracy of top-ranked results in generative recommendation.  

\subsubsection{Online A/B Test}
To empirically assess the practical business impact of $\textit{R}^\textit{3}$\textit{-VAE}, we further execute an online A/B test. The experiment evaluated user engagement using two primary metrics: StayTime/U, defined as the average total time spent by a user per day, and LongStay/U, which measures the count of documents on which a user's stay time exceeds ten seconds per user.

Table~\ref{tab:industrial_online_results} shows the online performance. The results demonstrate that $\textit{R}^\textit{3}$\textit{-VAE} drives significant improvements in user engagement metrics. For StayTime/U, $\textit{R}^\textit{3}$\textit{-VAE} achieves 2672s, which is 1.60\% higher than RQ-VAE (2630s) and 0.83\% higher than R-KMeans (2650s). 
Regarding LongStay/U, $\textit{R}^\textit{3}$\textit{-VAE} yields the highest score 25.35, surpassing RQ-VAE (25.20) and R-KMeans (25.28) by 0.60\% and 0.28\%, respectively.
These online results fully validate that the superior offline performance of $\textit{R}^\textit{3}$\textit{-VAE} translates into tangible business value, effectively enhancing user engagement in real industrial recommendation scenarios.  

\subsection{Discriminative Recommendation Performance on Industrial Dataset}
To validate $\textit{R}^\textit{3}$\textit{-VAE}’s generalization ability in discriminative recommendation scenarios (like Click-Through Rate prediction, a core task in industrial recommender systems), we conduct experiments by replacing the id embedding with each SID on the same industrial dataset.
\subsubsection{Implementation Details}
The experiments are conducted on the CTR model. The codebook size is $8192\times8192\times8192$. The framework follows DIN~\cite{zhou2018deep}, SIM~\cite{pi2020search}, and RankMixer~\cite{zhu2025rankmixer}. We use RMSPropV2 as the optimizer, set batch size to 2048 and initial learning rate to $1 \times 10^{-3}$, and leverage the recent 60 user click sequences as historical behavior features. 
\subsubsection{Offline Evaluation}
We adopt User Area Under ROC Curve (UAUC) as the core offline evaluation metric. Table~\ref{tab:metrics2} presents the offline UAUC performance and SID quality metrics of $\textit{R}^\textit{3}$\textit{-VAE} and baselines. It can be observed that $\textit{R}^\textit{3}$\textit{-VAE} achieves the highest UAUC (0.6669) across all methods, which aligns with its optimal SID quality (highest SC=0.94, lowest PD=-1.980, lowest CR=1.033, and smallest Gini=0.0314).
\subsubsection{Online A/B Test}
To quantify the actual gains of the $\textit{R}^\textit{3}$\textit{-VAE} for online services, we conduct an A/B test of RQ-VAE, R-KMeans and $\textit{R}^\textit{3}$\textit{-VAE}, with StayDuration/U (Average Stay Duration per User) and Cold-Start Click Volume as key metrics for Existing and New Users. 
Table~\ref{tab:industrial_online_results_din} presents the results of the A/B test. The data demonstrate that the $\textit{R}^\textit{3}$\textit{-VAE} achieves a 0.11\% and 0.65\% improvement in the StayDuration/U for existing and new users compared to R-KMeans, respectively. Meanwhile, the model also drives a significant increase in the click volume of cold-start content (with impression counts below 512), with respective lifts of 4.26\% and 15.36\% observed among existing and new users. These findings indicate that high-quality SID can bring substantial gains to industrial recommendation systems. Currently, our proposed $\textit{R}^\textit{3}$\textit{-VAE} model has been fully deployed in production services.

\begin{table}
\centering
\caption{Generative offline MRR performance and online A/B test results on industrial dataset. The \textbf{bold} values indicate the best performance, and the \underline{underlined} values denote the second-best performance.}
\begin{tabular}{l|lll}
\hline
    
\hline

\hline
Method & MRR $\uparrow$          & StayTime/U $\uparrow$ & LongStay/U $\uparrow$ \\
\hline
RQ-VAE    & 0.605        & 2630    & 25.20      \\
R-KMeans     & \underline{0.618}    & \underline{2650}    & \underline{25.28}     \\
$\textbf{R}^\textbf{3}$\textbf{-VAE} & \textbf{0.628~~~~(+1.62\%)}& \textbf{2672~~~~(+0.83\%)} & \textbf{25.35~~~~(+0.28\%)}  \\
\hline
    
\hline

\hline
\end{tabular}
\label{tab:industrial_online_results}
% \vspace{-4.5mm}
\end{table}
\begin{table}
    \centering
    \caption{Downstream discriminative recommendation performance and SID quality metrics of $\textit{R}^\textit{3}$\textit{-VAE} and baselines on industrial dataset.}
    \begin{tabular}{l|c|c|c|c|c}
    \hline
    
    \hline
    
    \hline
        \multicolumn{1}{c}{Method} & 
        \multicolumn{1}{c}{UAUC $\uparrow$} & 
        \multicolumn{1}{c}{Semantic $\uparrow$} &
        \multicolumn{1}{c}{Preference $\downarrow$} &
        \multicolumn{1}{c}{Collision $\downarrow$} &
        \multicolumn{1}{c}{Gini $\downarrow$} \\
        \multicolumn{1}{c}{} & 
        \multicolumn{1}{c}{} & 
        \multicolumn{1}{c}{Cohesion} &
        \multicolumn{1}{c}{Discrimination} &
        \multicolumn{1}{c}{Rate} &
        \multicolumn{1}{c}{Coefficient} \\
        \hline
        RQ-VAE&0.6538&0.76&-1.950&1.271&0.0815 \\ 
        OPQ-KMeans&0.6571&0.85&-1.975&1.070&0.0692 \\ 
        R-KMeans&0.6577&0.92&-1.985&1.039&0.0366 \\  
        MQ&0.6553&0.91&-1.960&1.037&0.0348  \\  
        $\textbf{R}^\textbf{3}$\textbf{-VAE}&\textbf{0.6669}&\textbf{0.94}&\textbf{-1.980}&\textbf{1.033}&\textbf{0.0314} \\ 
    \hline
    
    \hline
    
    \hline
    \end{tabular}
    \label{tab:metrics2}
\end{table}
\begin{table}
\centering
\caption{Discriminative online A/B test result on the industrial dataset. The \textbf{bold} values indicate the best performance, and the \underline{underlined} values denote the second-best performance.SD/U means StayDuration/U. CS Click means Cold-Start Click Volume.}
\begin{tabular}{l|cc|cc}
\hline
    
\hline

\hline
\multicolumn{1}{c}{Method} & 
\multicolumn{2}{c}{Existing Users} & 
\multicolumn{2}{c}{New Users} \\
\cline{2-3} \cline{4-5}
\multicolumn{1}{c}{} & 
\multicolumn{1}{c}{SD/U $\uparrow$} & 
\multicolumn{1}{c}{CS Click $\uparrow$} & 
\multicolumn{1}{c}{SD/U $\uparrow$} & 
\multicolumn{1}{c}{CS Click $\uparrow$}  \\
\hline
RQ-VAE    & 37680        & 8224    & 5039 & 1441      \\
R-KMeans     & \underline{37696}    & \underline{8335}    & \underline{5057} & \underline{1510}     \\
$\textbf{R}^\textbf{3}$\textbf{-VAE} & \textbf{37737 (+0.11\%)}& \textbf{8690 (+4.26\%)} & \textbf{5090 (+0.65\%)} &\textbf{1742 (+15.36\%)}  \\
\hline
    
\hline

\hline
\end{tabular}
\label{tab:industrial_online_results_din}
\end{table}
\begin{table}
    \centering
    \caption{Downstream GR performance and SID quality metrics of $\textit{R}^\textit{3}$\textit{-VAE} and baselines on Beauty.}
    \begin{tabular}{l|c|c|c|c|c}
    \hline
    
    \hline
    
    \hline
        \multicolumn{1}{c}{Method} & 
        \multicolumn{1}{c}{Recall @ 10 $\uparrow$} & 
        \multicolumn{1}{c}{Semantic $\uparrow$} &
        \multicolumn{1}{c}{Preference $\downarrow$} &
        \multicolumn{1}{c}{Collision $\downarrow$} &
        \multicolumn{1}{c}{Gini $\downarrow$} \\
        \multicolumn{1}{c}{} & 
        \multicolumn{1}{c}{} & 
        \multicolumn{1}{c}{Cohesion} &
        \multicolumn{1}{c}{Discrimination} &
        \multicolumn{1}{c}{Rate} &
        \multicolumn{1}{c}{Coefficient} \\
        \hline
        VQ-VAE&0.0552&0.85&-0.17&60.5&0.43 \\  
        RQ-VAE&0.0593&0.93&-1.18&48.99&0.89 \\ 
        KMeans&0.0599&0.93&-0.22&47.27&0.49 \\ 
        OPQ-KMeans&0.0595&0.94&-1.14&4.98&0.68 \\ 
        R-KMeans&0.0639&0.96&-1.39&1.14&0.11 \\  
        MQ&0.0563&0.90&-1.03&89.64&0.93  \\  
        $\textbf{R}^\textbf{3}$\textbf{-VAE}&\textbf{0.0716}&\textbf{0.97}&\textbf{-1.81}&\textbf{1.11}&\textbf{0.09} \\ 
    \hline
    
    \hline
    
    \hline
    \end{tabular}
    \label{tab:metrics}
\end{table}

\subsection{Correlation Between SID Metrics and Downstream Performance}
To further explore the relationship between SID quality metrics and downstream performance, we conduct a Spearman rank correlation analysis between the SID metrics (our proposed Semantic Cohesion, Preference Discrimination, and Collision Rate~\cite{li2025semantic}, Gini coefficient~\cite{fu2025forge}) and UAUC (the core downstream discriminative recommendation indicator) or Recall@10 (the core downstream GR indicator) in Table~\ref{tab:metrics2}. The Spearman correlation coefficient quantifies the monotonic relationship between two variables, with values ranging from -1 (perfect negative correlation) to 1 (perfect positive correlation).

\begin{table}
    \centering
    \caption{Spearman rank correlation between SID metrics and UAUC or Recall@10.}
    \begin{tabular}{l|cc}
    \hline
    
    \hline
    
    \hline
        \multicolumn{1}{c}{Metric} & 
        \multicolumn{1}{c}{Spearman Correlation} & 
        \multicolumn{1}{c}{Spearman Correlation} \\
        \multicolumn{1}{c}{} & 
        \multicolumn{1}{c}{with UAUC} & 
        \multicolumn{1}{c}{with Recall@10} \\
        \hline
        Semantic Cohesion &0.90&0.94 \\
        Preference Discrimination &-0.90&-0.75 \\
        Collision Rate&-0.70&-0.93 \\
        Gini Coefficient&-0.70&-0.64 \\
    \hline
    
    \hline
    
    \hline
    \end{tabular}
    \label{tab:spearman2}
\end{table}
\begin{table*}
    \centering
    \caption{Ablation studies of key components on Amazon-Beauty.}
    \begin{tabular}{l|c|c|c|c|c|c|c|c}
    \hline
    
    \hline
    
    \hline
        \multicolumn{1}{c}{Method} & 
        \multicolumn{3}{c}{Recall $\uparrow$} & 
        \multicolumn{3}{c}{NDCG $\uparrow$} & 
        \multicolumn{1}{c}{SC $\uparrow$} & 
        \multicolumn{1}{c}{PD $\downarrow$} \\
        \cline{2-4} \cline{5-7}
        \multicolumn{1}{c}{} & 
        \multicolumn{1}{c}{@10} & 
        \multicolumn{1}{c}{@20} & 
        \multicolumn{1}{c}{@30} & 
        \multicolumn{1}{c}{@10} & 
        \multicolumn{1}{c}{@20} & 
        \multicolumn{1}{c}{@30} & 
        \multicolumn{1}{c}{} & 
        \multicolumn{1}{c}{} \\
        \hline
        $\textbf{R}^\textbf{3}$\textbf{-VAE~(Full)}&\textbf{0.0716}&\textbf{0.1025}&\textbf{0.1250}&\textbf{0.0393}&\textbf{0.0470}&\textbf{0.0518}&\textbf{0.97}&\textbf{-1.81} \\
        w/o Reference Vector&0.0642&0.0975&0.1211&0.0357&0.0437&0.0486&0.94&-1.73 \\ 
        w/o Rating Quantization&0.0640&0.0968&0.1201&0.0351&0.0434&0.0478&0.90&-1.78\\ 
        w/o SC Regularization&0.0652&0.0988&0.1217&0.0369&0.0451&0.0490&0.90&-1.85 \\ 
        w/o PD Regularization&0.0647&0.0981&0.1214&0.0367&0.0440&0.0487&0.96&-1.56 \\
    \hline
    
    \hline
    
    \hline
    \end{tabular}
    \label{tab:ablate}
    \vspace{3mm}
\end{table*}

From the results in Table~\ref{tab:spearman2}, we observe consistent yet scenario-adaptive correlation patterns between our proposed SID quality metrics and downstream recommendation performance across both industrial discriminative recommendation tasks (UAUC) and public dataset generative retrieval tasks (Recall@10). Specifically, Semantic Cohesion (SC) demonstrates a strong positive correlation with both UAUC (\(\rho = 0.90\)) and Recall@10 (\(\rho = 0.94\)). This confirms that higher intra-cluster semantic consistency of SID universally enhances recommendation accuracy, which aligns with $\textit{R}^\textit{3}$\textit{-VAE} achieving the top SC values and leading downstream performance across all scenarios.
Preference Discrimination (PD) exhibits strong negative correlations with both metrics (\(\rho = -0.90\) for UAUC; \(\rho = -0.75\) for Recall@10). A more negative PD value indicates larger inter-cluster preference divergence, which helps distinguish user preferences and thus improves recommendation relevance.
Collision Rate (CR) shows a strong negative correlation with Recall@10 (\(\rho = -0.93\)) and a moderate negative correlation with UAUC (\(\rho = -0.70\)). Lower CR values mean fewer SID collisions (i.e., distinct items map to unique SID), reducing retrieval ambiguity and boosting performance—an effect that is more pronounced on public generative retrieval tasks. The Gini coefficient (a measure of codebook imbalance) presents moderate negative correlations with both UAUC (\(\rho = -0.70\)) and Recall@10 (\(\rho = -0.64\)), suggesting it has a relatively limited impact on downstream performance compared with SC, PD, and CR.

These unified results reveal that SC and PD are more effective SID quality proxies for industrial discriminative recommendation tasks, while SC and CR show stronger relevance to generative retrieval performance on public datasets. However, considering that PD can be optimized directly via end-to-end training, whereas CR typically requires additional post-processing strategies (e.g., codebook balancing strategy), we ultimately select SC and PD as the core optimization objectives and SID quality evaluation metrics for $\textit{R}^\textit{3}$\textit{-VAE} from the perspective of engineering feasibility. Notably, $\textit{R}^\textit{3}$\textit{-VAE} achieves the optimal combination of these key metrics across both industrial and public datasets, which explains its consistent superiority over baselines.

\subsection{Ablation Studies}
\subsubsection{Impact of key components}
To isolate the contribution of each core component of $\textit{R}^\textit{3}$\textit{-VAE}, we evaluate several ablated variants on Beauty dataset:  
\begin{itemize}
    \item $\textit{R}^\textit{3}$\textit{-VAE} w/o Reference Vector: Removes the reference vector projection layer; initial residual $\mathbf{e}^{(0)} = \mathbf{x}$.
    \item $\textit{R}^\textit{3}$\textit{-VAE} w/o Rating Quantization: Replaces the dot product rating mechanism with STE (same as RQ-VAE’s gradient approximation).
    \item $\textit{R}^\textit{3}$\textit{-VAE} w/o SC/PD Regularization: Removes the metric-aware regularization term.  
\end{itemize}

Table~\ref{tab:ablate} shows the results. Removing the reference vector projection layer reduces Recall@10 by 10.3\% and NDCG@10 by 9.2\%, confirming its role in mitigating initialization sensitivity and aligning residuals with semantic structure.  
Replacing the rating quantization module with STE reduces NDCG@10 by 10.6\%, verifying the rating mechanism’s value for gradient preservation. Removing SC/PD regularization also weakens the downstream GR performance, highlighting the metrics' role in guiding SID quality.  

\begin{table}
    \centering
    \caption{Ablation studies of regularization weight ($\lambda$).}
    \begin{tabular}{l|c|c|c|c}
    \hline
    
    \hline
    
    \hline
        \multicolumn{1}{c}{$\lambda$} & 
        \multicolumn{1}{c}{Recall @ 10 $\uparrow$} & 
        \multicolumn{1}{c}{NDCG @ 10 $\uparrow$} & 
        \multicolumn{1}{c}{Semantic $\uparrow$} &
        \multicolumn{1}{c}{Preference $\downarrow$}\\
        \multicolumn{1}{c}{} & 
        \multicolumn{1}{c}{} & 
        \multicolumn{1}{c}{} & 
        \multicolumn{1}{c}{Cohesion} &
        \multicolumn{1}{c}{Discrimination}\\
        \hline
        0.001&0.0656&0.0371&0.96&-1.79\\ %0.0440&0.0291
        0.01&0.0716&0.0393&0.97&-1.81 \\ %0.0445&0.0301
        0.1&0.0669&0.0373&0.95&-1.87 \\ %0.0457&0.0308
        % 1&0.0443&0.0303&0.95&-1.82 \\ %0.0644 0.0368
        % 10&0.0415&0.0277&0.96&-1.83 \\ %0.0628 0.0345
    \hline
    
    \hline
    
    \hline
    \end{tabular}
    \label{tab:lambda}
    % \vspace{-5mm}
\end{table}
\subsubsection{Regularization Weight}
Table~\ref{tab:lambda} shows Recall@10, NDCG@10, SC and PD as $\lambda$ increases from 0.001 to 0.1 (with $L=3$, $K=256$ fixed). $\lambda=0.01$ is the optimal choice in this scenario.  

\begin{figure}[htbp]
    \centering
    \vspace{-2mm}
    \includegraphics[width=\linewidth]{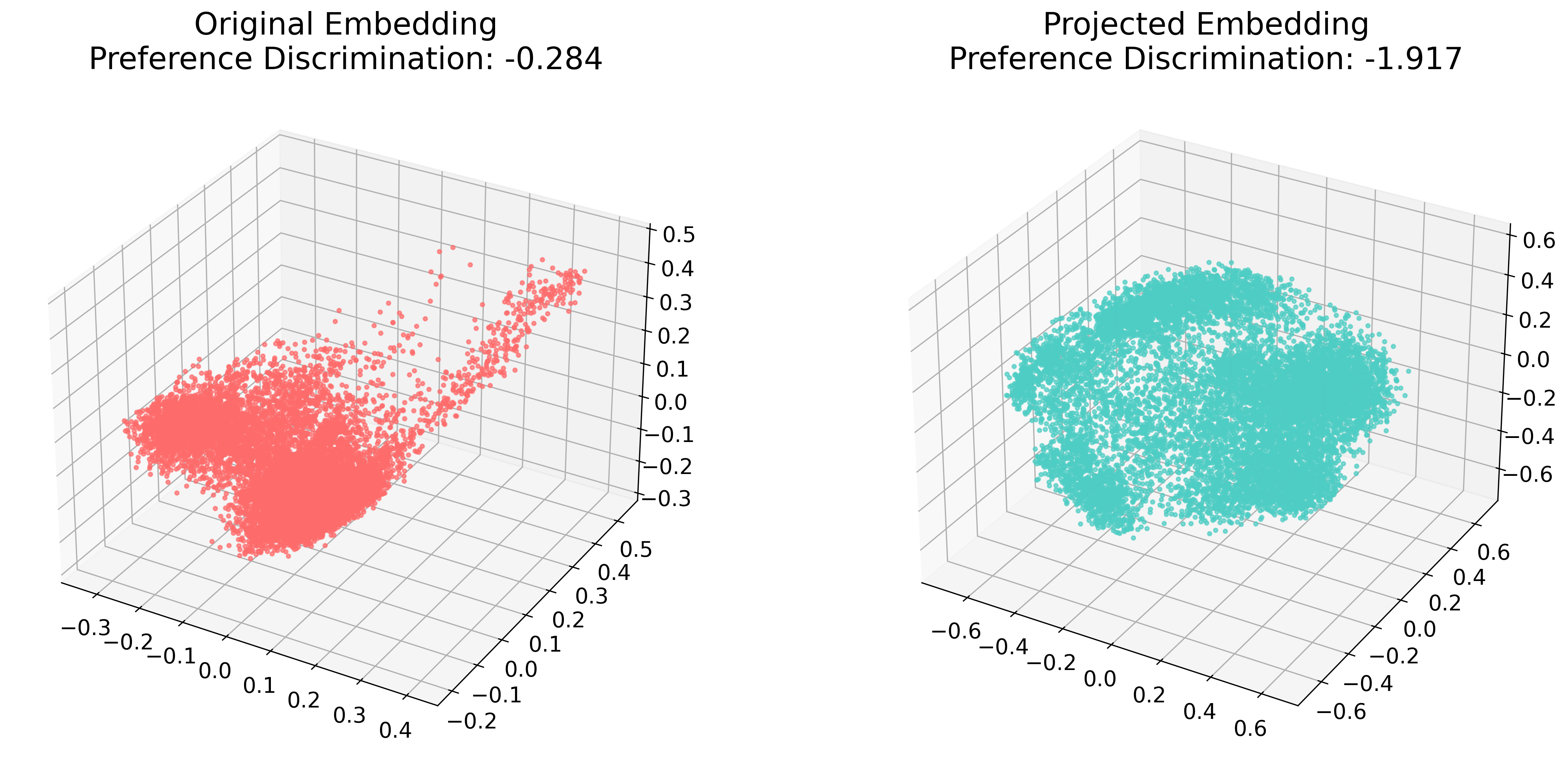}
    \vspace{-3mm}
    \caption{3D PCA visualization of embeddings before (left) and after (right) reference vector projection. The projected embeddings exhibit a more dispersed distribution, improving cluster separability.}
    \label{fig:3d_pca_vis}
    \vspace{-3mm}
\end{figure}
\begin{figure}[htbp]
% \vspace{-2mm}
    \centering
    \includegraphics[width=\linewidth]{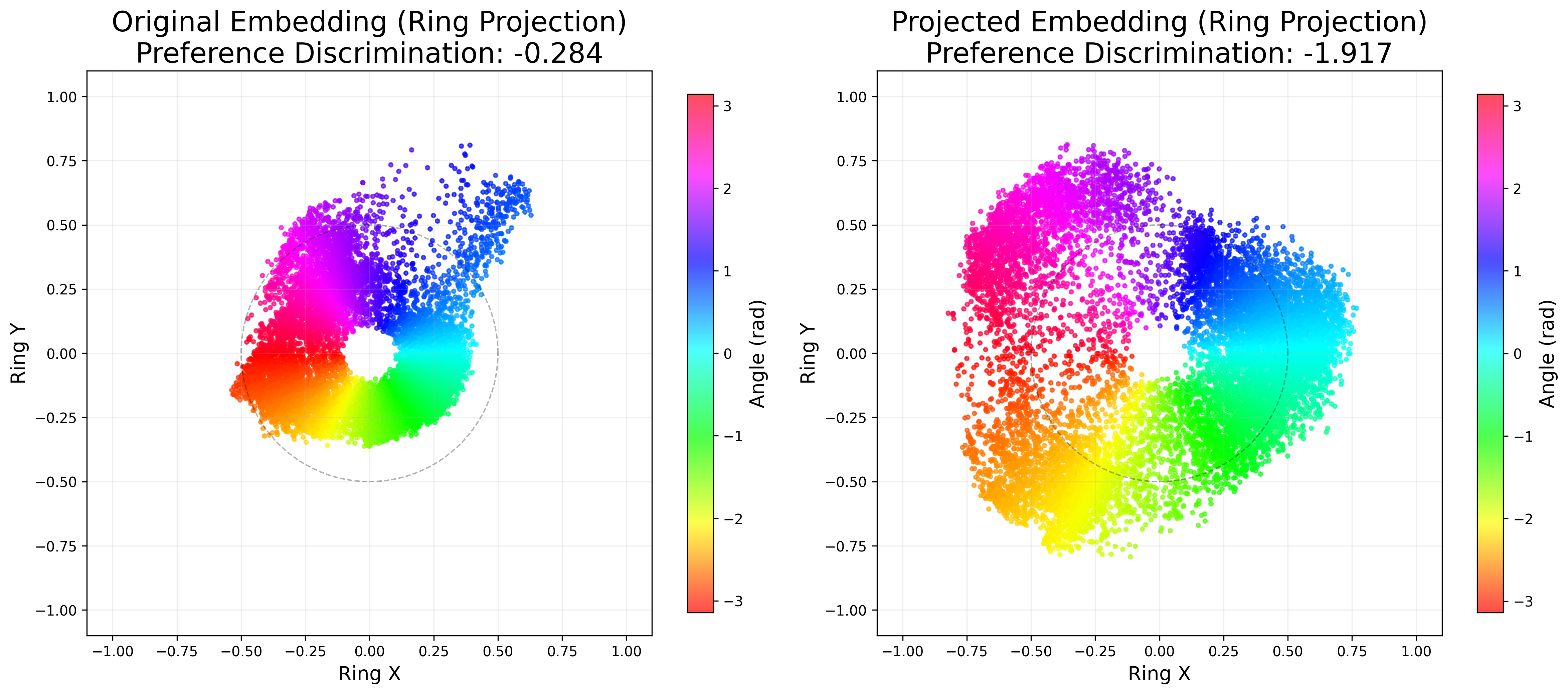}
    \vspace{-2mm}
    \caption{2D ring projection of embeddings (colored by angular position) before (left) and after (right) projection. The projected embeddings achieve a more uniform angular distribution (lower preference discrimination score), enhancing cluster distinguishability.}
    \label{fig:2d_ring_vis}
    \vspace{-2mm}
\end{figure}
\begin{figure}[t]
\vspace{-2mm}
\centering
\includegraphics[width=\linewidth]{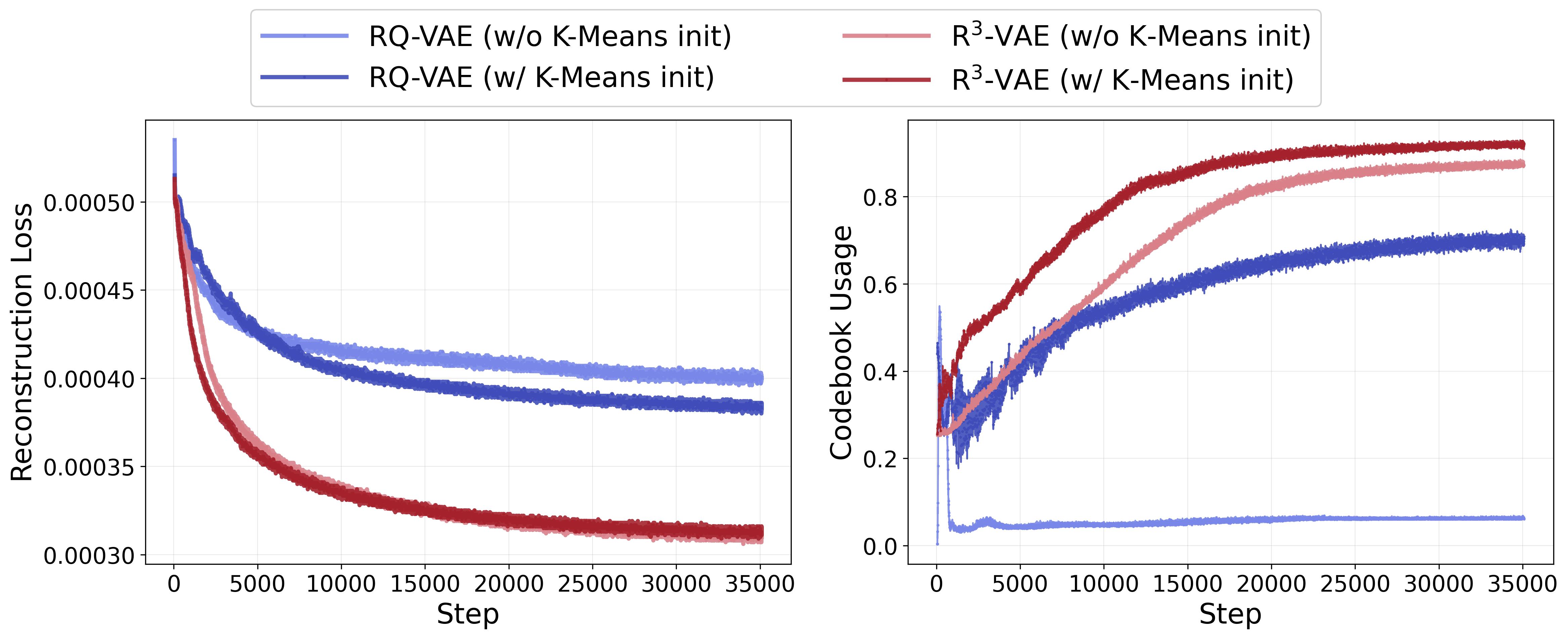}
\vspace{-2mm}
\caption{Training stability comparison: reconstruction loss (left) and codebook usage (right) of $\textit{R}^\textit{3}$\textit{-VAE} and RQ-VAE (with/without KMeans initialization).}
\label{fig:training_stability}
\vspace{-2mm}
\end{figure}

\subsection{Qualitative Analysis}
To further validate the effectiveness of the reference vector projection layer in shaping embedding distributions for clustering, we visualize the embeddings before and after projection using two methods: Spherical Distribution (Figure~\ref{fig:3d_pca_vis}) and Preference Discrimination Angle Distribution (Figure~\ref{fig:2d_ring_vis}).

\subsubsection{Visualization Analysis}
\paragraph{Spherical Distribution}
Figure~\ref{fig:3d_pca_vis} compares the 3D PCA projections of original embeddings (left) and projected embeddings (right). The original embeddings exhibit a compact, clustered distribution (concentrated in a small subregion of the PCA space), which limits the discriminability of clusters—similar points are overly aggregated, making it difficult for clustering algorithms to distinguish distinct groups. In contrast, the projected embeddings are spatially dispersed (spread across a broader range of the PCA space) while maintaining structural coherence. This dispersion aligns with the design goal of the reference vector projection layer: it expands the embedding space to enhance inter-cluster separability, a critical prerequisite for effective clustering.

\paragraph{Preference Discrimination Angle Distribution}
To quantify the distribution of preference discrimination, we use a 2D ring projection (Figure~\ref{fig:2d_ring_vis}), where points are colored by their angular position (rad) to reflect the distribution spread. For the original embeddings (left), the preference discrimination score is \(-0.284\)—the points are highly concentrated in a narrow angular range (e.g., red/yellow regions dominate), leading to severe overlap. After projection (right), the preference discrimination score drops to \(-1.917\), and points are evenly distributed across all angular regions (covering red, yellow, green, blue, and purple zones). This dispersion ensures that embeddings of distinct clusters occupy non-overlapping subspaces, directly facilitating accurate clustering.

The visualizations in Figure~\ref{fig:3d_pca_vis} and Figure~\ref{fig:2d_ring_vis} collectively demonstrate that the reference vector projection layer transforms embeddings from a compact, overlapping distribution to a dispersed and clustering-friendly distribution. By expanding the embedding space and enhancing inter-cluster separability, this layer lays a robust foundation for downstream clustering tasks—addressing the core limitation of original embeddings (poor discriminability) and enabling more accurate and stable cluster assignments.

\subsection{Training Stability Analysis}
\label{sec:train_stable}
To further validate the training stability of $\textit{R}^\textit{3}$\textit{-VAE} (a key advantage of its gradient-preserving rating mechanism) and its insensitivity to initialization strategies, we compare the reconstruction loss and codebook usage curves of $\textit{R}^\textit{3}$\textit{-VAE} and RQ-VAE (with/without K-Means initialization) during training, as shown in Fig. \ref{fig:training_stability}.

\subsubsection{Reconstruction Loss Behavior}
The left subfigure depicts the reconstruction loss curve over training steps. It can be observed that $\textit{R}^\textit{3}$\textit{-VAE} (both with/without K-Means initialization) shows consistently fast convergence and a tight loss gap between initialization strategies. In contrast, RQ-VAE exhibits a large performance discrepancy between initialization settings. This demonstrates that $\textit{R}^\textit{3}$\textit{-VAE}’s training process is robust to initialization choices, avoiding the sensitivity that plagues traditional RQ methods. RQ-VAE (even with K-Means initialization) shows higher loss fluctuation and a final loss (0.00038) that is 19\% higher than $\textit{R}^\textit{3}$\textit{-VAE}. 

\subsubsection{Codebook Usage Behavior}
The right subfigure presents the codebook usage (proportion of active codewords) over training steps. $\textit{R}^\textit{3}$\textit{-VAE} achieves near-full codebook usage (approaching 1.0) after ~20,000 steps, and maintains stable usage throughout training. RQ-VAE with K-Means initialization only reaches ~0.7 codebook usage, while RQ-VAE without K-Means initialization suffers from severe codebook collapse (usage drops to ~0.05 and remains stagnant). 
$\textit{R}^\textit{3}$\textit{-VAE} exhibits superior training stability in both loss convergence and codebook utilization, regardless of initialization strategy. This robustness stems from its core innovations (gradient-preserving rating mechanism and reference vector anchoring), which address the training instability bottlenecks of existing VQ-based methods.

\section{Conclusion}
In this paper, we address two critical limitations of existing Semantic Identifier (SID) generation methods for generative recommendation: training instability caused by inadequate gradient propagation and inefficient SID evaluation that relies on costly downstream validation. We propose a novel framework named \textbf{Reference Vector-Guided Rating Residual Quantization VAE ($\textbf{R}^\textbf{3}$-VAE)} alongside two SID metrics: Semantic Cohesion and Preference Discrimination. Both metrics exhibit strong correlations with GR performance. Empirical results on six benchmarks demonstrate that $\textit{R}^\textit{3}$\textit{-VAE} outperforms state-of-the-art methods, achieving an average improvement of 14.5\% in Recall@10 and 15.5\% in NDCG@10 across three public datasets (Beauty, Sports, and Toys). Experiments of the GR and CTR model on a large-scale industrial dataset from Toutiao further demonstrate that our new method exhibits excellent applicability and business value in real-world industrial recommendation scenarios.

% \clearpage
\bibliographystyle{plainnat}
\bibliography{main}

\clearpage
\section{Contributions and Acknowledgments}
\label{contributions}

% All authors of R3-VAE are listed in order of their contribution to the work.

\begin{multicols}{2} %
\sffamily{\color{seedblue}  \large{Team Toutiao Recommendation}} \\
\\
\color{seedblue}Qiang Wan\\
\color{seedblue}Ze Yang\\
\color{seedblue}Dawei Yang\\
\color{seedblue}Ying Fan\\
\color{seedblue}Xin Yan\\
\color{seedblue}Siyang Liu\\
\\
\\
\\
\color{seedblue}Yicong Liu\\
\color{seedblue}Chenwei Zhang\\
\color{seedblue}Wei Xu\\
\color{seedblue}Jiahao Qin\\
\color{seedblue}Ke Wang\\
% \color{seedblue}Pixun Li \\
\end{multicols}

% \clearpage
% \beginappendix
% \input{sections/appendix}

\end{document}